# Skini: Reactive Programming for Interactive Structured Music

Bertrand Petit[a] and Manuel Serrano[a]

a   Inria, Sophia-Antipolis, France

**Abstract**

    This paper presents Skini, a programming methodology and an execution environment for interactive structured music. With this system, the composer *programs* his scores in the HipHop.js synchronous reactive language. They are then executed, or *played*, in live concerts, in interaction with the audience. The system aims at helping composers to find a good balance between the determinism of the compositions and the nondeterminism of the interactions with the public. Each execution of a Skini score yields to a different but aesthetically consistent interpretation.

    This work raises many questions in the musical fields. *How to combine composition and interaction? How to control the musical style when the audience influences what is to play next? What are the possible connections with generative music?* These are important questions for the Skini system but they are out of the scope of this paper that focuses exclusively on the computer science aspects of the system. From that perspective, the main questions are *how to program the scores* and *in which language?* General purpose languages are inappropriate because their elementary constructs (*i.e.,* variables, functions, loops, etc.) do not match the constructions needed to express music and musical constraints. We show that synchronous programming languages are a much better fit because they rely on temporal constructs that can be directly used to represent musical scores and because their malleability enables composers to experiment easily with artistic variations of their initial scores.

    The paper mostly focuses on scores programming. It exposes the process a composer should follow from his very first musical intuitions up to the generation of a musical artifact. The paper presents some excerpts of the programming of a classical music composition that it then precisely relates to an actual recording. Examples of techno music and jazz are also presented, with audio artifact, to demonstrate the versatility of the system. Finally, brief presentations of past live concerts are presented as an evidence of viability of the system.

# The Art, Science, and Engineering of Programming

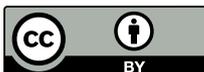

## 1 Introduction

This paper presents a composition method based on synchronous programming, and a music production environment named Skini, for performing collaborative music in live concerts. Skini music is collaborative in that the audience participation impacts the score that is played either by musicians or by computer programs. The goal of Skini, is to evaluate the relevance of the implementation of a composition method based on the HipHop.js language. After several experiments, we have chosen to base Skini music on three key concepts:

1. The interaction with the audience governs the music to be played next. The music starts and evolves only upon audience requests. This continuous interaction, that is implemented as web interfaces running on audience smartphones, impacts the evolution of the musical piece being executed.

2. The music is made of sequences of *patterns* that are elementary musical elements, such as short (few seconds) musical sequences, or mere sounds. Patterns are used to implement the interaction with the audience. At each moment of the show, the audience is presented a set of patterns that they can select. Selecting patterns enables or disables other patterns for the future.

3. Although interactive, Skini music follows a strict structure that has been created by a composer beforehand. Different executions of the same composition with different audiences will deliver different musical outcomes but they will all be artistically consistent as they will all reflect the artistic orientations of the composer. Two executions will be different as they will depend on different non deterministic interactions with the audience but they will still be two instances of the same composition and recognizable as such, as could be different interpretations of the same standard by a jazz groups (see for instance, the countless versions of "My Favorite Things" recorded by saxophonist John Coltrane that are all different but still all recognizable).

Skini music seeks for a balance between composition and improvisation as jazz music does, but with major difference: with Skini, the improvisation is due to the audience and not to the musicians. The musical structure is defined by the composer, which enforces the musical coherence of the piece, the music execution is executed by musicians or by computer programs, the improvisation is under the responsibility of the audience.

A traditional occidental musical score is a two-dimensional description of music on a staff notation. The horizontal axis describes the course of time and the duration of notes. The vertical axis describes the pitch of the notes. Skini scores are by all means different. It does not describe music on the standard staff notation. The purpose of this article is to explain the Skini score structure that we have experimented with. The paper sketches how the music composer elaborates a piece with Skini and how it translates his artistic creation into a computer program that is executed during the show. The paper will be illustrated with musical artifacts and the description of actual concerts that took place during the past two years.

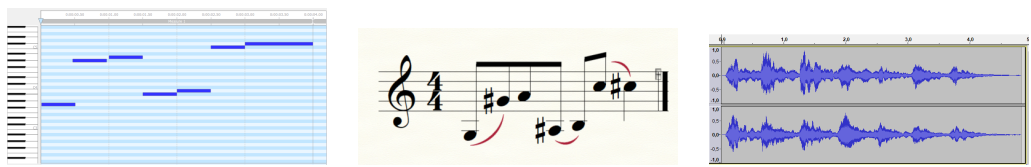

■ **Figure 1** A pattern is a short musical sequence. It can be implemented as a MIDI sequence (left), a score (middle), a sound wave (right), or any other format that enables sound production.

The organization of the paper is as follows. Section 2 presents the artistic creation process promoted by Skini. Section 3 presents how the musical scores are transformed into computer programs. Section 4 presents the HipHop.js programming language used to program the scores. Section 5 shows the actual score programming. Section 7 briefly presents the Skini runtime environment. Section 8 presents three musical pieces implemented and played with Skini. Section 9 describes the actual live concerts that have been played with the system. Sections 10 and 11 present the related work and conclude the paper.

## 2    The Artistic Process

A musical work in the sense of Skini consists of an assembly, driven by an audience in real time, of basic musical elements that we call *patterns*, according to a predefined organization that we call *musical paths*. The basic elements and the predefined organization are designed by a composer. These components constituting the materials of the work do not have an effective existence until they are in the hands of an audience. We could make the connection with a score and a musician. The score is a material that is music only if a musician performs it.

### 2.1   Patterns

A Skini piece of music is a form of *synthesis by concatenation* of patterns. That is to say that the music produced will be the result of the sucession and the superimposition of patterns. Patterns can be mere sounds, short recorded or synthesized music sequences, or complete or partial musical phrases, see figure 1. The audience selects them during an interpretation. The order and the actual elements that are selected at any moment during the show create a unique interpretation of the musical composition.

### 2.2   Interaction

Music is produced by playing patterns according to *audience selections*. We call *selection* a request generated by an audience member who wants a particular pattern to be played. Audience members are informed in real time of the list of available patterns that they can listen to on their handsets. This poses several problems:

- Generally there will be more requests for patterns than patterns available. It is a question of managing the excess of requests in relation to pattern resources;
- By choosing to associate patterns with instruments, we are confronted with the fact that a given instrument can only play one pattern at a time. A musician only plays one score. To play several patterns, it will therefore be necessary to provide them sequentially to an instrument or a musician so that he can play them one after the other;
- We want to synchronize the pattern starts to ensure musical coherence;
- We want to assure everyone that their selection will have an impact on the performance.

Among the possible solutions for these points, such as abandoning certain patterns, mixing patterns, introducing voting mechanisms, we chose to implement a queuing mechanism for each instrument, because it was the method that best addressed all of our concerns. This mechanism involves finding a way to inform the audience about the waiting times generated by queues.

Other problems appear:

- The interaction requires the composer to be well aware of the impact of pattern durations on the likely behavior of the audience. If the patterns are too long, for a too large audience, and not enough instruments, the audience can be demotivated by too long waiting times. If the patterns are too short, the audience will have difficulty identifying them.
- The system has to control the number of patterns that a member of the audience can select in succession in order to prevent the monopolization of the system by particularly active or malicious individuals. That is, the number of patterns a member can send to the queues before they have been played. In our productions we limit these requests to no more than three.
- The interaction may involve audience requests that are not selections of patterns, but choices that allow voting mechanisms to decide on the overall evolution of the music, for example.

Controlling the interaction and giving structure to the musical pieces is the definition of Musical Paths that we will see now.

## 2.3 Scores and Musical Paths

Once the patterns have been created, the composer organizes them into *groups* and *tanks*. Patterns are accessible for *selection* to the audience only via groups and tanks. We call activating a group the act of *making it available for selection by the audience.* Patterns that belong to a group can be selected as long as the group is active. Patterns that belong to a tank can only be selected once as long as the tank is active. The *musical path* is the sequencing of group and tank activations and de-activations.

A group is activated or de-activated based on the audience interaction. For instance, the composer can plan to activate a group of patterns only after a fixed number of patterns of another group have been selected. The group activation/de-activation also depends on constraints between groups. For instance, the composer can use exclusion

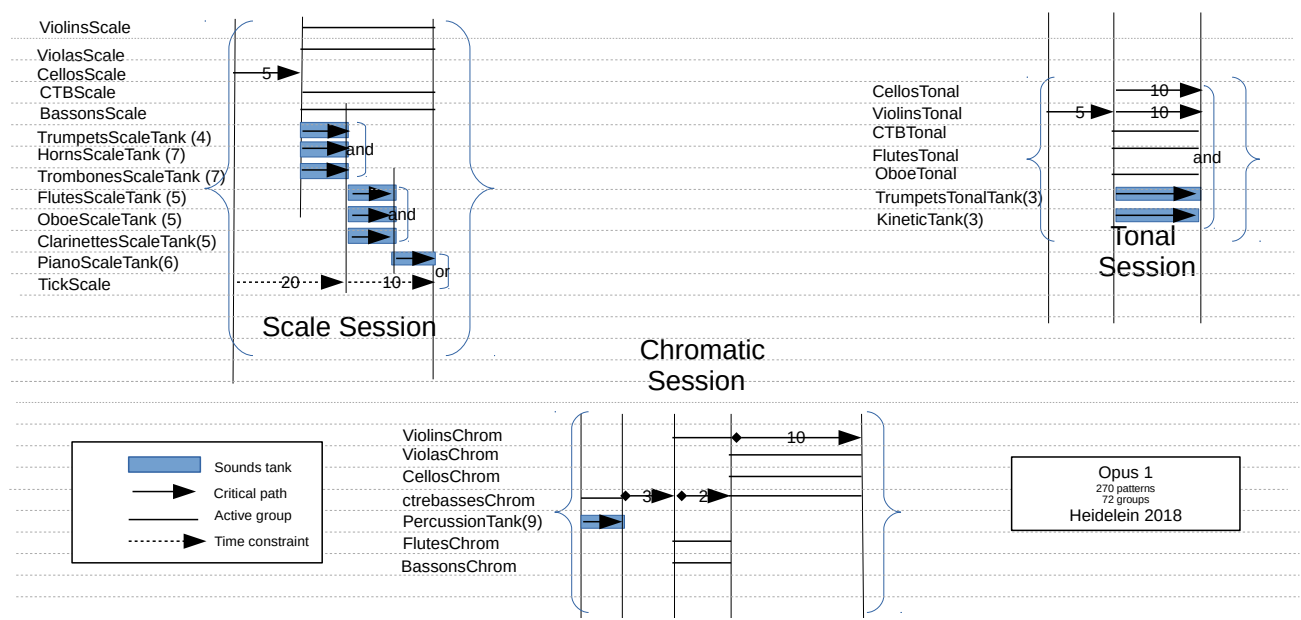

■ **Figure 2**   A graphical representation of Opus1

rules between groups that contain incompatible instruments, or he can impose a group after another in order to avoid too repetitive selections of the audience. We have two kinds of groups, *repetitive groups* (groups for short) and *tanks*. Patterns of repetitive groups can be selected without limit as long as the group is active. Tank patterns are disabled when they are selected. Tanks are a way to avoid repetitions for example. The organization of groups, that is the first step toward the actual programming of the score, can be sketched graphically. The diagram in figure 2 is a description of a musical piece. Numbers in between brackets next to the group names show how many patterns are contained in the tanks (for instance the TrumpetScaleTank contains 4 patterns). It makes explicit the dependencies between groups using various graphical conventions of our owns. For instance, the bracket labeled with an "and" of the Scale session means that the three tanks (trumpets, horns, and trombones) must all be empty, that is all the patterns of the tank must have been selected by the audience, for the next groups, flute, oboe, and clarinets, to be activated. For the groups, a labeled arrow means that the audience must select at least that number of patterns before the next step is executed. For instance, at least 5 patterns of the CelloScale group must be selected before the patterns of the groups can be selected by the audience. An arrow without a label means that the patterns of the group can be selected as long as the whole group is active.

This drawing is an example of a first step of the artistic process. It abstracts away many details and logical and time constraints that are too difficult to express graphically. The next step is, of course, to transpose the initial artistic ideas into a computer program that can be executed during actual concerts.

## 3   From Scores to Programs

The sketch of figure 2 can be used by the composer to define independent modules the piece should contain. In what precise order the groups should be activated and de-activated? How long the activation of a group should last? How many pattern selections from one group should activate a successor? Which groups are incompatible with others? Etc. All these questions can be addressed more or less precisely before writing the computer program.

As already apparent in figure 2 the program implementing the score will have to simulate a large state machine, where the states will correspond to group activation and de-activation and the transitions between the states will model the audience interaction and the passing time. The complexity of the automata obviously depends on the complexity of the score and on the complexity of the musical orchestration. We have observed that scores are usually complex. For instance, The Opus1 piece involves 270 patterns that are assembled in about 70 different groups. The underlying automata that encodes the constraints between the groups counts about 10000 states! The obvious question raised by this complexity is, *which programming language to choose to implement these programs?*

Event-based programs are known to be difficult to elaborate with traditional general programming languages [3, 9, 16] because of the underlying automata and their possibly exponential number of states that have to be programmed explicitly, generally by the means of global variables, mutations, and functions. The problem becomes even more critical when the system evolves as any modification of the specification generally requires a new automaton. With Skini, because of the trial and error mechanism inherent in musical creation, programs will evolve a lot! Composers will probably initiate their program with a musical idea, they will then make some experiments, and work on many variations. As the program is the product of this artistic creation, there cannot be any fixed specification known in advance! General programming languages are thus inappropriate.

Specialized musical languages such as MAX/MSP or PureData are widely used in computer music. They are mostly meant for programming musical streams and although they can be used to program automata, only relatively mildly complex problems are in reach. We have found that reactive programming languages (Esterel [23], Lustre [7], and Signal [12]) that rely on high-level orchestration operators are much better adapted. They enable composers to reason in terms of *parallel execution*, *repetition*, *synchronization*, and *preemption* that directly map the constraints of the scores. This makes it efficient to explore variations of an orchestration. This is precisely what is needed for the process of the artistic creation of musical orchestration. The underlying automata controlling the music is kept implicit and never exposed to the programmer. As the web is the classical Skini deployment platform, we have opted for the HipHop.js programming language that is a dialect of Esterel for JavaScript. In the next section we introduce that language and in section 5, we show how we use it to program musical scores.

## 4   HɪᴘHᴏᴘ.ᴊѕ

HɪᴘHᴏᴘ.ᴊѕ is a synchronous reactive language, originally designed for the web and IoT, that adds synchronous concurrency and preemption to Hᴏᴘ.ᴊѕ [26], which is itself a multitier extension of JavaScript. HɪᴘHᴏᴘ.ᴊѕ simplifies the programming of non-trivial temporal behaviors that are at the heart of musical scores.

HɪᴘHᴏᴘ.ᴊѕ [4] is an evolution of the HipHop [5] programming language that has been originally designed for the Scheme programming language. After several endeavors [31] HɪᴘHᴏᴘ.ᴊѕ has now reached its final shape. This last evolution of the language differs significantly from its ancestors. We present it briefly in this section, only presenting its main concepts and constructs that are essential for this paper. This section does not assume any prior knowledge to synchronous reactive programming.

Unlike Hᴏᴘ.ᴊѕ and JavaScript, which mostly deal with asynchronous communication between distributed components, HɪᴘHᴏᴘ.ᴊѕ belongs to the class of synchronous languages that were introduced in the 1980's, initially to program embedded systems for which orchestrating activities is a major concern. There key design feature is the reaction to external events in a conceptually instantaneous and deterministic way: examples are real-time process control, graphical user interfaces, communication protocols interfaces, and, more generally, state machine-based designs. In this work, it is used for more than that: it is also the language used to express and execute pattern-based musical scores. In other words, a Sᴋɪɴɪ musical score is an HɪᴘHᴏᴘ.ᴊѕ program and giving an HɪᴘHᴏᴘ.ᴊѕ concert is running this HɪᴘHᴏᴘ.ᴊѕ program, with interaction with the audience, of course.

JavaScript is based on an atomic execution mode, thus implementing a kind of synchronous reaction. Along its evolution, it has been augmented by useful features for dealing with asynchronous orchestration: `yield`, which blocks the control and restarts from there at next execution; `async`, that enables a sequential programming style for asynchronous computations, and promises that simplify the use of `async`. These statements which facilitate writing basic temporal behaviors in JavaScript are at the heart of the HɪᴘHᴏᴘ.ᴊѕ design, but the language offers many more high-level statements that add much more programming power: *temporal[1] statements*, that are statements which process signals or a signal expression, such as emitting a signal with a value, waiting for a signal or or a signal and a specific value of this signal, etc; *synchronous concurrency*, which is fundamental to modularize designs and make them easily reusable; *instantaneous signaling* to coordinate concurrent executions using broadcast signals; and *preemption* that makes it possible to control the lifetime of active temporal statements, starting them, suspending them, and killing them when needed. The conjunction of these statements makes the behaviour of complex executions explicitly visible, which is, in our view, essential to reduce the complexity of the elaboration of music scores.

---

[1] We will meet this term frequently. In synchronous programming terminology, "temporal" is not defined in relation to "time", but in relation to "signal states", that are not limited to time, as this word might suggest. This is an abbreviation for "signal status at the moment JavaScript causes a reaction of the HɪᴘHᴏᴘ.ᴊѕ program".

## 4.1  Principles of HipHop.js

A HipHop.js program executes steps called *reactions* or *instants*. Internally, steps execute HipHop.js statements in sequence or in parallel, letting them communicate using *broadcast signals*. At each of these reactions, each signal broadcasts a unique *present / absent status* to all statements in the same scope, and optionally a unique *value* belonging to an arbitrary data type. If needed, the value can be initialized within the HipHop.js signal declaration. The run-time interface between HipHop.js and JavaScript is realized by a JavaScript *reactive machine*, that provides a simple communication API between them.

A JavaScript program activates a reactive machine by calling a react() function associated to that machine. The machine acts in JavaScript via a "standard" mechanism of listener.

Signals are divided into *input signals* whose status and values are set by the host JavaScript program and passed to HipHop.js via the reactive machine, *output signals* that are returned to JavaScript via the reactive machine when the reaction terminates, and *local signals* used internally by the HipHop.js program and invisible from the reactive machine.

By default, the broadcast status of a signal S is *absent* and the optional value remains that of the previous reaction. The signal can be set *present* for the reaction if it is an input set present by the reactive machine or if an "emit S()" statement is executed by the HipHop.js program, in both cases optionally with a new value. Its current status is returned by the JavaScript expression "S.now", and its current value by the JavaScript expression "S.nowval". These expressions can be used wherever a JavaScript expression can appear in an HipHop.js module, *i.e.*, in data computations and tests. They play a critical role for triggering the HipHop.js temporal statements that are essential for the high-level handling of complex temporal control.

Conceptually, each reaction is instantaneous with a unique status and value for each signal, which are seen by all HipHop.js statements executed in the reaction, the new ones if the signal just got emitted. This is called the *synchrony hypothesis* [2]. It ensures that concurrency is *deterministic*, unlike for asynchronous concurrency. To realize the synchrony hypothesis in practice, a HipHop.js program is compiled into a standard JavaScript code run by the reactive machine that executes as usual until no progress can be made, synchronous concurrency being achieved by micro scheduling of sequential and concurrent statements. The fact that the generated code is sequential and executed atomically avoids any run-time interference between the HipHop.js execution and the JavaScript environment.

## 4.2  Control Flow

Let us consider the module JazzQuartet simulating a Jazz group where the rhythmic players will be continuously playing and two soloists, a sax and a trumpet will alternate. This could be simulated with the following program:

```
1  hiphop module JazzQuartet( in song, in solo ) {
2    fork {
```

```
3      run Drums( ... );
4    } par {
5      run Bass( ... );
6    } par {
7      every( solo.now ) {
8        if( solo.nowval === "sax" ) {
9          run Sax( ... );
10       } else if( solo.nowval === "trumpet" ) {
11         run Trumpet( ... );
12       }
13     }
14   }
15 }
```

The JazzQuartet module involves four submodules: Drums, Bass, Sax, and Trumpet. Each submodule is in charge of playing the eponymous instrument. It takes as input the song event that tells each instrument which song to play, and the solo event that controls which soloist, sax or trumpet should play at any moment.

The run construct (line 3, 5, 9, and 10) instantiates a submodule. Running a module means inlining its body at the location where it is invoked and binding its environment signals in the lexical scope of the caller. The form "run M(...)" means that each interface signal is implicitly bound to a signal of the same name in the lexical environment, if there is one. This is a way to avoid declaring each of the signals used by a module, when it is invoked by a module that already uses the same signals. In our example in line 9, the Sax module will use the signals from the JazzQuartet module. Note that '...' is a keyword of the language and not an ellipsis with use for writing this paper. Manual binding of differently named signals is done by "S1 as S2". This is a way to rename a signal when it is used under a new name in another module. The name of the signal in the called module S1 is associated with the signal of the calling module S2. This syntax is not used in this example.

The fork/par construct runs its branches in parallel. Operationally, it picks one branch that executes until it can make no further progress because it is blocked on a test for the status or value of some yet unknown signal, then another branch executes in the same way, etc. Because a branch can emit a signal another branch is waiting for, it may happen that executing one branch enables another one to resume, and so on. This micro scheduling is invisible to the user. It is realized at runtime in the current implementation, in a way which makes the result scheduling-independent and thus deterministic.

The every construct (lines 7) implements a preemptive loop that checks for a condition. In this example, each time a solo event is emitted, the associated value is checked for preempting the current soloist and selecting the another one. An every loop starts its body when its condition is true, executes its body, but kills it immediately and restarts it afresh each time the condition is met anew in some further instant. It is called a *strongly preemptive* statement: at any instant where the statement is alive and the condition is true, the current instance of the body *does not execute*, and the body is immediately restarted afresh. The condition can be any JavaScript expression evaluated in a lexical environment where all the visible signals are qualified by now and, nowval.

Finally, note that the if operator used line 8 is also part of the HipHop.js language, in spite of its similar with its JavaScript counterpart. Its expression test uses the same syntax as the every construct.

### 4.3 Events

In the previous example, the expression solo.now is true if and only if the signal solo is emitted in the instant. When that condition is met, the every preempts its body and one of the Sax or Trumpet module is executed. If one instance was already running, this one is first killed. When the select module terminates, the every loop simply waits for a new emission of that signal.

A signal is emitted from outside the program by the hosting JavaScript program that controls the machine reaction or it can be emitted from within HipHop.js using the "emit signame( optValue )" instruction.

Execution threads can synchronize on events. This is the purpose of the await construct. The instruction "await( expr )" blocks the current thread until expr is satisfied. Synchrony and broadcasting imply that all expressions of the whole HipHop.js program evaluate in the same signal environment. That is, if two different expressions use the same signal and evaluate during the same instant, they will see the same value for that signal, no matter their execution order. For instance, considering the following statement:

```
1  fork {
2    if( sax.nowval > 10 ) ...
3  } par {
4    emit sax( 30 );
5  } par {
6    if( sax.nowval > 20 ) ...
7  }
```

Here, a micro scheduling of substatements in the instant may start by evaluating line 2, block there because sax is yet unknown, then line 4 that emits sax, line 6, and line 2 again, which is now executable since sax is known to be present. Both line 2 and line 6 will see the value 30 for the expression sax.nowval. Any other valid scheduling would give the same result. Enforcing this property requires a dependency analysis that tells for each expression when it is ready to evaluate. In this example, it tells that the tests lines 2 and 6 cannot be executed before line 4. When deadlocks occur, they are always detected but reported at runtime.

### 4.4 Asynchronous Actions

Controlling the instruments that play the music is a primordial aspect of the music execution. With Skini, HipHop.js is in charge of controlling when an instrument starts, stops, or changes its configuration, but the actual playing made by the instrument is left to an external controller. An instrument can be controlled by another electronic system itself controlled by HᴏᴘHᴏᴘ.ᴊs, or it can be played by a musician that inter- prets the instructions it receives from HᴏᴘHᴏᴘ.ᴊs. From the HᴏᴘHᴏᴘ.ᴊs programming perspective, whatever the actual implementation of the instruments, they are imple- mented as asynchronous tasks that interact with the rest of the HᴏᴘHᴏᴘ.ᴊs program by exchanging signals. Similarly, other components essential to the music execution will be implemented as external background tasks. For instance, for implementing the rhythmic session, a background clock pulsing at regular physical intervals is needed. This can be implemented as follows:

```
1  hiphop module Metronome( pulse ) {
2    async {
3      this.react( {[pulse.signame]: this.sec = 0} );
4      this.intv = setInterval( () => this.react( {[pulse.signame]: ++this.sec} ), 1000 ) );
5    } kill {
6      clearInterval( this.intv );
7    }
8  }
```

This module is a mere wrapping of the native asynchronous JavaScript setInterval function, but with two important aspects to notice. First, at each second generated by setInterval, it calls the reactive machine to create a new HᴏᴘHᴏᴘ.ᴊs instant with input the current value of pulse in seconds. Second, the kill optional form in async (line 6) tells that the resource used by the JavaScript setInterval function is freed as soon as the JavaScript async statement is killed for any reason, for instance, because the song is over. At lines 3, 4, and 6 this refers to the async block. The form "this.react()" provokes a new reaction of the reactive machine.

Using Metronome, we can now revise our jazz quartet example to prevent a soloist to play more than 30 consecutive seconds.

```
1  hiphop module JazzQuartetPreempt( in song, in solo ) {
2    signal pulse;
3
4    fork {
5      run JazzQuartet( ... );
6    } par {
7      run Metronome( ... );
8    } par {
9      every( pulse.nowval % 30 === 0 ) {
10       emit solo( "" );
11     }
12   }
13 }
```

This new module uses a local signal declaration (line 2). This is a signal that can only be used inside the machine. The module JazzQuartetPreempt merely runs in parallel the already existing JazzQuartet, with a metronome, and a guard that every 30 seconds stops the current solo player, if any.

This example, in addition to illustrating how to combined HᴏᴘHᴏᴘ.ᴊs determinism synchronous events and outside world asynchronous tasks, shows how HᴏᴘHᴏᴘ.ᴊs programs can be elaborated by combining simpler modules. In this example, the original JazzQuartet module is re-used *totally unchanged* in a more complex setting. This ability to compose by the means of high-level operators (parallelism, sequencing, and preemption components) more complex ones is one of the main HIPHOP.JS strength for music composition.

## 5  Programming Scores

In this Section we show how to use HIPHOP.JS to program musical scores with the SKINI platform. For that we use Opus1 example, the contemporary classical music piece that we have introduced in its early stage (see figure 2). More precisely, we show how to transform what was a mere early mockup into an executable HIPHOP.JS program that will let the composer play the music and modify it in an iterative process until he reaches the final aesthetic result.

A score program is almost a direct mapping from the graphical temporal diagram (see figure 2) into HIPHOP.JS constructs. The general principles of the implementation are as follows.

- Groups are implemented by two signals: one input signal and one output signal. The input signal is emitted when a participant of the audience selects a pattern of that group. The signal carries a value about the very pattern that has been selected. The output signal is emitted by the SKINI engine to activate or deactivate a group.
- Tanks are implemented as arrays of groups.
- Parallel arrows of the diagram are implemented as HIPHOP.JS fork/park constructs.
- Labeled arrows are implemented as await instructions.
- Connections between two arrows are implemented as a HIPHOP.JS sequence.

For simplicity, some technical details of the HIPHOP.JS programming are left unexplained. We focus on the elements that are mandatory to understand the general principles of the musical score programming.

Opus1 consists of three different successive musical moment. So, the main module of its implementation consists of three HIPHOP.JS submodules that are executed in sequence.

```
1  hiphop module Opus1() implements ScaleIntf, ChromaticIntf, TonalIntf {
2    run ScaleSession(...);
3    run ChromaticSession(...);
4    run TonalSession(...);
5  }
```

The Opus1 module (line 1) implements three interfaces. In HIPHOP.JS an interface is a list of signal declarations. Declaring that a module implements an interface is equivalent to declaring the interface signals in the module itself. The three run commands are executed sequentially. That is, line 3, the ChromaticSession module starts when ScaleSession is over and line 4, TonalSession starts when ChromaticSession is done executing. This synchronization is automatically implemented by the HIPHOP.JS sequence operator without any explicit mention of the program.

■ **Listing 1** The implementation of the Chromatic session.

```
1  hiphop module ChromaticSession() implements ChromaticIntf {
2    fork {
3      emit ChromBassOut(true);  // activate the bass group
4    } par {
5      run Tank(sigarray=ChromPercuTank); // activate the percussion tank
6    }
7    await count(3, ChromBassIn.now); // wait for the audience to select 3 bass
8    fork {
9      emit ChromViolinsOut(true); // activate the violins group
10   } par {
11     await count(2, ChromBassIn.now);
12   } par {
13     emit ChromFlutesOut(true); // activate the flutes group
14   } par {
15     emit ChromBassonsOut(true); // activate the bassons group
16   }
17   emit ChromFlutesOut(false); // deactivate the flutes group
18   emit ChromBassonsOut(false); // deactivate the flutes group
19   fork {
20     await count(10, ViolinsChromIn.now); // wait for 10 violins selections
21   } par {
22     emit ChromViolasOut(true); // activate the violins group
23   } par {
24     emit ChromCellosOut(true); // activate the violins group
25   }
26 }
```

Let us detail the implementation of the ChromaticSession module presented listing 1. Line 3, the emission of the ChromBassOut signal with value true activates that group. Simultaneously the tank of percussion instruments is started. This relies on the instantaneous semantics of the HipHop.js emit instruction. After the tank complete, line 7, three ChromBass audience pattern selections are waited. The meaning of an instruction "await count( N, expr )" is to wait of N occurrences of expr. This construct is used extensively in score programming. Immediately after the three bass selections happen the second movement of that session starts with the activation of the ChromViolins, ChromFlutes, and ChromBassons. The rest of the implementation follows the same principles. After module ChromaticSession completes, all groups must be deactivated by emitting all the output signals with false. To keep the source code as short as possible, this is not shown here.

ChromaticSession may seem overly simple to implement as it only involves trivial synchronizations. We will see with the ScaleSession that sometime more complex scheduling is needed but first let us present the implementation of the Tank module.

```
1  hiphop module Tank( var sigarray ) {
2    fork ${
3      sigarray.map( sig =>
4        hiphop {
5          await ($(sig + "In").now);
```

```
6          emit ${sig + "Out"}(false);
7       })
8   }
9 }
```

The Tank module is part of the Skini library and composers are not required and not even expected to write this sort of components. This module uses the multi-staging facilities of HipHop.js. A HipHop.js program is actually a JavaScript value that is constructed before being loaded inside the reactive machine. Line 1, the Tank module receives in the variable sigarray, a JavaScript array that contains all the input and output signals of the tanks. Line 2, the fork elaborates its body using the compile-time statement ${...}. For each elements of the array a HipHop.js sequence is built. It waits for the audience to select the group. As soon as this happen, it immediately deactivates the corresponding group.

## 6    Programming Experience

The HipHop.js program we have presented in the previous section enjoys two benefits over an hypothetical implementation in a general purpose language. First, its distance from the initial diagrams (see figure 2) of the composer is small. Although the mapping is not automatic, it is almost straightforward. Second, as the implementation relies on high level orchestration operators, it is not cluttered with many implementation details that would make it complex and difficult to change. The HipHop.js implementation is malleable. It is easy to try new variations and to observe the aesthetic impact of the modifications.

As pieces were composed, we noticed that the language allowed musical combinations that the composer had not foreseen with his graphical representation. Examples are superimpositions of entire parts of the score, controlling the stopping, restarting or aborting a session from another (with the abort and suspend commands of HipHop.js). Etc. We have also realized that if it easy to start with a graphic view, it is also relevant for an advanced composer, accustomed to HipHop.js, to follow the opposite approach. That is, to reason about the orchestration using HipHop.js constructs, and then make a graphic synthesis of it, allowing easier access to a musical analysis. The design of the score in HipHop.js allows logical combinations between signals that are impossible to express simply with a graph. There are therefore clearly several possible levels in the use of HipHop.js. It can be a graphic transcription in the lineage of "classic" scores, where HipHop.js is an implementation tool. But it can also be a way to managing complex combinations of events, outside the usual ways of writing music. In this case HipHop.js becomes a musical composition tool itself. However, it is not easy to program in HipHop.js as soon as the automatons become very complex. This is particularly the case when the composer wishes to introduce relationships between "sessions", i. e. HipHop.js modules, which will have to interact, generate, or destroy themselves. The implementation of a powerful debugger is part of the projects around HipHop.js to facilitate this type of development.

## 6.1 Simulation and Rehearsals

In our musical context, the programming experience does not only consist in writing a program but also in evaluating the musical result and modifying the program accordingly. To evaluate the music the composer needs to listen to it. This is the purpose of rehearsals. Unfortunately, rehearsal is a major obstacle to interactive music. As the audience is only available for the actual show, rehearsal is almost impossible. To work around this difficulty, we have developed a *simulator* whose objective is to behave randomly as a group of individuals. It simulates an audience of barely interested people that would just click randomly on the buttons of their user interfaces. A real audience is unlikely to push buttons completely at random, but as simple as it is, this tool has proven to be essential for the composer. The musical artifacts we refer to in this paper have been obtained by simulation.

The simulator has three main parameters: the minimum response time of people in the audience, a maximum response time, and the waiting time for a pattern to be played. Of course the behaviour of an audience is not equivalent to that of the simulator. This has been observed during actual shows. The activity of an audience is more "reflective", less systematic in the selection of patterns. In real performances, we have observed pattern discovery phases where the activity decreases and then starts again. Moreover, choices proposed about the evolution of the music, the selection of instruments, the athmosphere of the music are not made at random by an actual audience. The size of the audience and the cultural level of the audience play an important role in these cases. The three pieces presented below in this paper were designed using the simulator, and the actual musical results with audience for this pieces, were in line with the musical project.[2]

## 6.2 Development Life Cycle

Composing a musical score with HipHop.js is different from programming a more traditional computer application, with a specification that is generally known before the programming starts. Here, there is no specification per se. The composer using HipHop.js oscillates between programming and simulating, in a logic of discovering solutions produced by complex combinations, that are difficult to imagine a priori. These quick 'back and forth' between listening and programming are made possible by HipHop.js. They would not have been possible by means of a general purpose programming language.

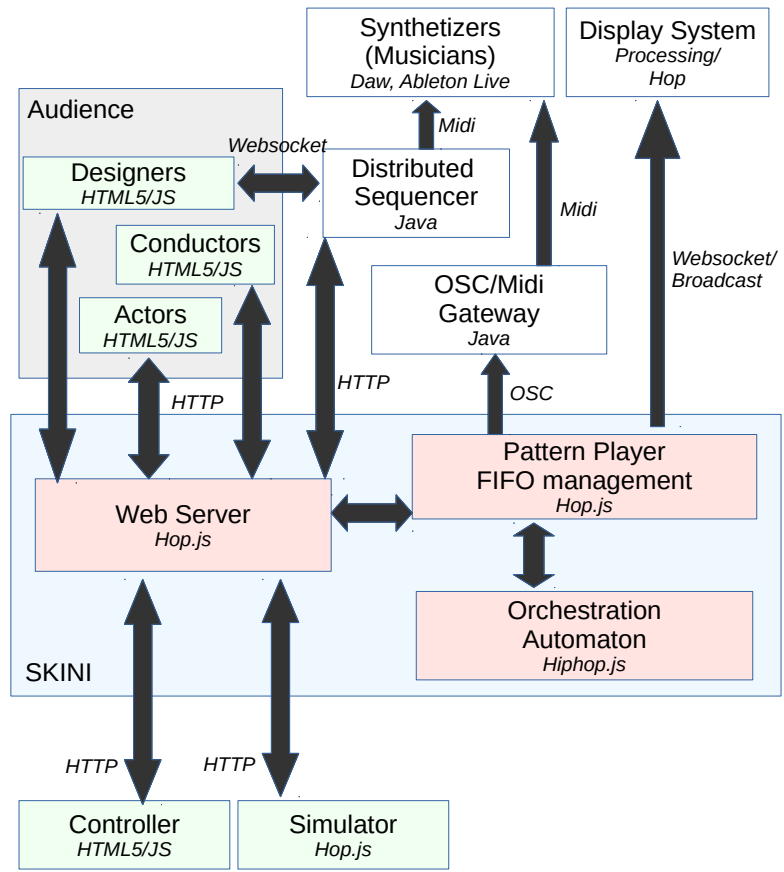

■ **Figure 3** Skini software architecture.

## 7 The Skini Execution Environment

Section 5 has presented the programming of music score as a transformation process that starts with the composer musical ideas and that ends with HipHop.js programs. This section overviews their execution environment.

The general Skini architecture is presented in figure 3. The central component is a web server that controls and supervises the whole application. It provides the single page web applications implementing the GUI used by the audience, for pattern selection. It controls the synthesizers that produce the actual music via an Open Sound Control (OSC)/MIDI gateway. It runs the score of the music composer (see section 5). It supervises a graphical display system that provides live feedback to the audience during a show. Two data structures are essential for the execution of a show.

---

[2] In addition to that, the simulator can be also considered as a mean to generate music automatically, which opens interesting applications in the domain of *generative music* [6].

- The *matrix of available groups of patterns*. It is controlled by the execution of the Hip Hop.js score and it tells at every moment of the show which groups and tanks are activated. The matrix is updated after Hip Hop.js output signal emissions. For instance, in the source code presented listing 1, the emission of the ChromBassOut will be intercepted by the JavaScript component running on the Skini server. It will also update the matrix by activating the chromatic ass group and it will trigger a refresh of the audience GUI.

- The *pattern queues*. As already mentioned in section 2.1, the pattern queues are fed by the audience and consumed by the synthesizers. When a participant of the audience selects a pattern, that is, when he clicks the associated button of the web interface, two actions are triggered. First, a signal is sent to the score. In reaction, this Hip Hop.js program updates the *matrix of available patterns*. For instance, after the third reception of the input signal ChromBassIn (listing 1, line 7), the Opus1 Hip Hop.js program enables the chords group and the trumpets, horns, and trombones tanks. Second, the MIDI sequence corresponding to the select pattern is pushed into a *pattern queue* (FIFO) that is consumed by the synthesizers.

As explained in section 2.2, the queues used by synthesizers are essential when the number of participants exceeds the number of available instruments. In such a configuration, at any given time, the server receives several requests to play several patterns for the same instrument. Since one instrument can only play one pattern at a time, not all the requested patterns can be played at once. To solve this problem, two strategies are possible: silently discarding the unsatisfied requests or pushing them into a queue. We have opted for the second solution because we think that it is more likely to preserve the audience's commitment to the show. By using queue, Skini avoids the frustration of someone activating a pattern without it being ever played. To maintain audience attention, the system also informs each participant of the expected delay before a selected pattern is played. Later, when that pattern is played a flash or a vibration is sent to the participant's phone, and also a message is included in the global display system. Being as responsive as possible for preserving the attention and focus of the audience is one of Skini priorities. We believe that if the ignoring some users requests might hinder the active participation of the audience to the show.

## 8  Three Pieces

In this Section we present three examples of pieces composed with Hip Hop.js. They are of different in styles to illustrate the Skini versatility. The scores are developed according to the programming principles presented in section 5. They are executed using the simulator described in section 7.

### 8.1  Jazz

The first piece is a jazz composition. It is available at https://soundcloud.com/user-651713160/grand-loup. It is performed by a set of "samplers" including a piano, percus-sion, bass, trumpet, saxophone, drums, and guitar. It uses 80 patterns organized in 7 groups. 8 patterns for piano, 6 for percussion, 8 for bass, 18 for trumpet, 18 for drums, 18 for saxophone and, 4 for guitar. These are small numbers for a Skini piece. The code is about 300 lines long. The timing and order of the patterns will vary depending on the audience's behavior or the simulator behavior. The presented recording has a simple sequence of scenes which starts with guitar and bass. The piano enters at second 7 and the percussion at second 15. From the second 18 we have a trumpet solo which is followed by a saxophone solo where the drums come on stage. The longest pattern of this piece is 4 seconds long.

This recording has been obtained by simulation. That is, the patterns are randomly activated but we have also played this piece on stage with a real audience. The outcome has been close to this simulation.

## 8.2 Electro

The second piece is for Electro music. It can be listened at https://soundcloud.com/user-651713160/lusine-mai-2019/s-gMdN9. It is relies on 10 synthesizers and is organized in 5 scenes, each controlled by a HipHop.js module similar to those presented section 5. The piece includes 102 patterns and 28 groups of patterns. It lasts about 10 minutes. The HipHop.js code is 280 lines long. It mainly controls sequences of activations, without complex behavior. The patterns were designed by young pupils during *Les Fabriques à Musique* described in section 9.2.

## 8.3 Opus1

Opus1 is the piece presented in figure 2. It is available at https://soundcloud.com/user-651713160/opus1-skini-v2-chrom. It uses 270 patterns organized in 73 groups of patterns, and 16 instruments (samplers). The overall HipHop.js source code is 460 lines long. The recorded piece lasts about 2:50 min. In the following, we detail how the music relates to the HipHop.js program. We focus on the Scale session that opens the piece and on its implementation that has been presented in listing 1, for which we present a chronological explanation. To help the reader, we have isolated that session in a dedicated music file available at: https://soundcloud.com/user-651713160/opus1-skini-v2-chrom.

0:00 The piece opens with bass that are activated by the emission of the signal ChromBassOut (line 3).

0:09 Just after the basses, we hear the percussion which were activated by the tank of line 5. We also can hear the violins activated in line 9, because the percussion tank was exhausted, and bass patterns are short. The await of line 7 was quickly taken into account.

0:21 The flutes, activated in line 13, enter after a aleatory timing defined by the stochastic behavior of the simulator (it could be an audience behavior).

0:42 The bassoons of line 15 enter, immediately followed by the violins if line 20.

0:52  From this point the piece evolves in a different manner than exposed in the HipHop.js code because the code example given here is in fact an extract of a longer piece.

## 9 Live Performances

Actual concerts and live performances have been played with Skini. *The Golem*, the first one, was given in 2017 in front of about 150 persons, in the context of the Nice *MANCA* contemporary music festival[3] Three other live shows took place in 2018 with smaller audiences of about 50 persons each. These were experimental concerts that allowed us to test new functionalities and new user interfaces, an important aspect that is out of the scope of the present paper. Aside the concerts, Skini is also used for educational projects. One, already completed, has been designed for introducing music composition to children of about twelve years old. It was funded by the SACEM for a program named *Fabrique à Musique*[4] This section briefly reports these experiments.

### 9.1 The Golem, MANCA 2017

The Golem show is based on the *legend of Golem* in Prague, who was created in the 16[th] century in order to protect the Jewish community. This performance is composed for a calligrapher/percussionist and an actress/singer. The Calame (pencils) of the calligrapher was used as a percussive instrument producing sounds that were processed in real-time. The show was executed with an early prototype of the Skini platform. The feedback we received from that experience has been essential to re-design and improve the system that has evolved to its present shape.

The show lasted 45 minutes and contained two interactive music scenes of about five minutes each. We ran the show twice: first, on December 8 2017, we ran a rehearsal with two groups of young pupils; second, on December 9, in front of an audience of 150 spectators in the context of the official festival. Before the show, some instructions were given to the audience on how to access the private WiFi network, how to get connected to the server, and the main features of the interface. The role of the public in the scenario of the *legend of the Golem* was also explained and emphasized. Even though it was an experiment of the early days, the two shows were successful and the participation to the interactive sessions were high considering the relative complexity of the interfaces that we were using back then. Amongst the 150 participants, 90 accessed the system and participated actively. The two sessions of interactions lasted 3 and 6 minutes. This generated 754 pattern activations, which makes roughly a mean of 1.4 sound events per second.

---

In this show HIPHOP.JS controlled the whole schedule of the performance, the sound effects for the percussion and the singer, and the launching of sound files and not only the interactive sessions.

This performance has been a strong incentive to pursue our efforts in the direction of the interaction concerts. This experiment and its successes and failures have been a highly valuable source of inspiration for improvements of the user interfaces and the way to compose interactive music.

### 9.2 Les Fabriques à Musique

*Les Fabriques à Musique* is a cultural program from SACEM to raise awareness among children and young people to musical creation, writing and composition. SKINI and a music composer were selected for the creation of a show by a class of 25 young pupils (12 years old). The goal was to create a complete musical work in 5 sessions of 2 hours. The work was presented officially in a concert hall of the Nice Conservatory of Music. This piece is presented in section 8.2.

The project consisted in assisting students to illustrate a story of their own with a musical composition. High school students designed the musical illustration with SKINI in groups of 5 or 6 children. This project was more than a mere performance; it was a complete educational approach. In this context, the SKINI platform was not only used for playing music created by a single composer, but also as a tool for collaborative music creation.

This was the first time SKINI was used as an educational tool. This project was well accepted by the school's music teacher, whose contribution to the smooth running of the various creative sessions was important. This positive reception was likely due to the fit between the tool and the pedagogy methodology in use in that school. The creative activities are emphasized, and children are very receptive to it. The children were highly committed to the project and had no problem in handling the tool. The playful aspect of the tool facilitated this creative process. However, a longer time period would have been needed to allow children to fully understand the notions of note length, scales, etc. This experiment encourages us to go further on the pedagogical dimension of the platform.

### 10    Related Work

Computer-Assisted Music (CAM) has a history that dates to the 1950s with the emergence of the first computers. Since then, research and development in this field have always closely followed the evolution of IT. Among the many fields dealing with computer music, SKINI is in line with collaborative music production systems. This chapter provides an overview of some significant achievements in this area.

As early as 1974, in the Philippines, Professor Jose Maceda set up a performance (Ugnayan) using radio sets [20]. The idea was to involve mass participants by broadcasting sounds on 37 radio channels in the city of Manila. Channels that everyone could select and broadcast in the street to create a collective sound, a form of collaborative sound collage. With the development of smartphones, interactive music initiatives have naturally increased tenfold. We have chosen a limited number of examples based on their relevance to our platform. Most of the articles cited here include states of the art.

*Dialtone: A Telesymphony* in 2001 is often cited as the first work that used smartphones for musical performance and demonstrated the potential of the mobile phone as an interface. Benjamin Taylor in his article "A History of the Audience as a Speaker Array" of 2017 [29] gives a detailed overview of the use of smartphones as a loudspeaker set. The idea of using smartphone speakers for performance has since been widely exploited. Ircam's recent COSIMA project [11] is part of this continuity by using the smartphone as a musical instrument and by integrating the latest developments in W3C's Web Audio technologies. The FAUST project of GRAME [19] or Kronos of the Center for Music and Technology of the Sibelius Academy in Finland [18] are also part of this trend by offering development tools to create instruments, including on smartphones. Skini is in line with these projects, but does not use smartphone speakers as a means of producing music, but as a "local" way of listening to patterns.

*Open Symphony* [34] is a recent (2017) interactive system based on web technologies, that allows the audience to control a score using voting mechanisms. The score is played by musicians. As with Open Symphony, Skini proposes to the audience to act as a "meta-composer/conductor". But unlike Open Symphony, Skini does not rely solely on a voting mechanism to activate and create the score being played.

The *massMobile platform* [33] is a client-server system developed for mass participation of an audience during a performance using smartphones. This project in its definition is very close to Skini. It has been used for large-scale performance. The global architecture is based on a Java server (SOAP and GlassFish) interfaced with MAX/MSP [24] via APIs developed for the project. This is a project that dates back to 2011, with an architecture that is in line with the technologies popular at that time. The system is able to control the graphical interface of smartphones via MAX/MSP. The global architecture is nevertheless the only thing that Skini shares with this system. Skini has a more precise musical project, and does not rely on MAX/MSP as a tool for musical creation. MAX/MSP is a tool with very rich functionalities, but at a low level. Skini, on the other hand, offers a fairly precise and high-level framework for musical creation. We will also see that Skini being more recent, is based on JavaScript for server and not Java.

*Echobo* [13] also proposes to use smartphones as musical instruments. It is the smartphone's speaker that produces the sounds. Echobo is based on the principle of a conductor who defines the musical structure at a high level (progression of chord, scale). From this structure the audience can play note by note, while being globally harmonically aligned. We are quite far from Skini's musical project, but there are similarities, particularly on the principle of the conductor. There is a major difference in sound production, Skini does not use smartphone speakers to produce music, but a central system.

*Open Band* [27] is an original collaborative system that proposes to use the alphabet to produce experimental music. Each participant can produce a message that is loaded into a server to be translated into a sequence of sounds played on the smartphone. A controller can act on the music produced by playing on the volume or audio frequencies for example. The music produced is created by associating a sound sample with each ASCII character, or using syntheses produced with Web Audio APIs. This system has some similarities with ours. It proposes a method of music production and not access to a low-level tool like massMobile does. It also includes a notion of controller. Nevertheless as in the case of echobo the sound is produced by smartphones unlike Skini which is closer to massMobile at this level. Moreover, the technique of music production is radically different from that proposed by Skini.

*Swarmed* [8] is a 2013 project. It is interesting on a historical basis. It places a strong emphasis on the technical architecture based on web technologies and a private Wifi network. Swarmed concurrently runs 7 centralized instruments that can be controlled by the audience. The sound is produced by the *Csound* software. Skini uses an architecture similar to that of Swarmed. There is a central server that controls a set of instruments with HTML clients. Unlike Skini, Swarmed does not offer composition logic. It provides a suite of instruments whose use by the audience is completely free.

Even if they are not totally in the same field of use as Skini, a number of collaborative tools on the market deserve attention, such as *Ableton Link* and online software such as Soundtrap [14], PiBox [21], Flat and TrackD [30].

Although *Live Coding*, with solutions such as superCollider [17], Chuck [32] or Sonic Pi [1], is also a field that deals with programming languages and music production, it is not directly connected to the production of music by an audience. Skini does not synthesize sound from code, but from pattern concatenation. Nevertheless, among the future evolutions we can imagine using interactions between Live Coding solutions for sound synthesis and Skini for control via the audience. Similarly, it would be quite possible to push developments in the direction of interaction with live programming tools such as TextAlive [10] or sketch-in-sketch [25] or to use a graphic representation system such as Threnoscope [15] as an interface with the audience.

## 11 Conclusion

Skini is the result of a research work on the implementation of an interactive music composition method that can guarantee coherent aesthetic results.

It proposes to extend the concept of musical score by accommodating new dimensions to the musical composition process. With Skini, the composer designs the music not only according to a predefined aesthetic but also according to the interaction he plans with the audience. We came up with a method of composition based on patterns, group of patterns, and management of different type of interactions that can be used to compose different styles of music, as demonstrated by the three examples we have commented (jazz, electro, and classical music). At the heart of the composition process are *scores*. These are computer programs that simulate complex state machines. This is known to be difficult to implement with general purpose languages. Instead we advocate the use of synchronous reactive programming languages because the high-level orchestration operators (parallelism, sequence, synchronization, preemption) they rely on are precisely those that are needed to program scores. We have seen that the score design can take place according to several scenarios depending on the composer's computer skills. It can follow the classic path from graphic specification to simulation, or go directly from idea to coding and simulation. HipHop.js proved to be a good tool in both cases. Coding in HipHop.js is better adapted than a general purpose language for our approach, and concrete achievements have demonstrated the relevance of our composition method. However, the difficulties inherent in our method should not be underestimated. Automata programming is certainly greatly facilitated by HipHop.js, but it remains a delicate problem and requires the acquisition of solid programming skills.

In this paper, we have presented the whole artistic process that starts with mere musical inspiration, and explain why it make sense to program scores for interactive music with an audience using the HipHop.js language, but Skini is likely to address other areas of music production such as music for video games for example [28], where interaction plays an important role. In the article [22] the author explicitly refers to interactive music as a technique for extending linearly composed music. The technology implemented by Skini is most likely applicable to music creation for video games thanks to its ability to take into account and efficiently manage a large number of variables related to a game interaction.

## About the authors

**Bertrand Petit** Is author of this article and member of the Indes team of Inria Sophia-Antipolis. Contact him at bertrand.petit@ inria.fr.

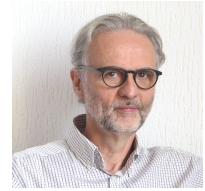

**Manuel Serrano** Is author of this article, Research Director and leader of the Indes Team at Inria Sophia-Antipolis. Contact him at manuel.serrano@inria.fr.

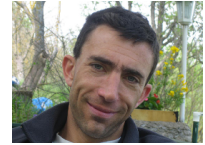